# HOW PARTICLES CAN EMERGE IN A RELATIVISTIC VERSION OF BOHMIAN QUANTUM FIELD THEORY: Part 2 - Fermions

## T. Mark Harder

*E-mail address:* tmharder1@shockers.wichita.edu

It is shown how Fermionic material particles can emerge from a covariant formulation of de Broglie-Bohm theory. Material particles are continuous fields, formed as the eigenvalue of the Schrodinger field operator, evaluated along a Bohmian trajectory. The motivation for this work is due to a theorem proved by Malament that states there cannot be a relativistic quantum mechanics of localizable particles.

Within the arena of Bohmian theory, there are many points of view that do not necessarily agree with each other. The principle point of view is that observables have values at all times, and uncertainties enter at the primary level because of entanglement of the system under observation with the apparatus used to observe it. One hears the phrase "particle *or* field becomes particle *and* field". The "particle" seemingly plays a minor role in the formalism, and is realized as the entity that is causing the observed disturbance. We would like to take a minimalistic view, and assert that the "particle" is actually already contained in the formalism, and is the object we are presenting here. The novelty of the Bohm theory in this viewpoint becomes the existence of the trajectory equations at a fundamental level.

In a continuation of this topic, it is shown how to fit Fermions into the description of material particles, using similar arguments presented in a preceding paper [1], with special considerations made for Fermionic fields. The arguments are similar, with care allowed for the handling of Grassmann valued functionals and methodology borrowed from reference [2]. There may be some unfamiliarity with the Schrodinger picture in Fermionic quantum field theory, and background material is included for those readers for whom this situation applies. It is hoped that this will also provide clarity to the arguments.

Let us state our position at the beginning. In part we observed that a single mode excitation of the vacuum, for arbitrary time dependence, provided a simple first order equation. Solving this equation provided a stationary solution, and then a Lorentz boost (the trajectory here is a rectilinear movement at constant speed) provides the final solution. This object was an eigenvalue of the Schrodinger operator, evaluated along a trajectory. A direct translation of this procedure to the Grassmann variables is not desirable. One can observe, however, that each component of the Dirac equation solves the $2^{nd}$ order wave equation, and apply the bosonic argument to this component. The result is still Grassmann valued, and so we "bosonize" the result (see [3]) to get real valued quantities. The philosophy behind this ansatz is commonly applied in scattering theory so that particle creation operators for Bosons and Fermions have the same form.

We make one final remark. Reference [4] presents and proves a no-go theorem on the localizability of "particles" in a quantum theory. Ascribing to the "particle" a set of reasonable mathematical properties (a Hilbert space representation, a strongly continuous unitary representation of the translation group, a unique self adjoint Hamiltonian and a localizability condition) the theorem states that for a theory satisfying these conditions the probability of finding the particle in any finite spatial set is zero. This theorem is the motivation for the exercise of this study.

## FORMALISM FOR FERMIONS IN SCHRODINGER PICTURE

In this section we will briefly review formalism taken from [2] and [5]. We will adopt the nomenclature of a two dimensional Fermionic harmonic oscillator, define the Schrodinger wave functionals and states and construct Gaussian states for the purpose of generalizing the notion of a Fock vacuum. This notation is necessary to cover the general case of charged spinors of both positive and negative energy. In addition, the additional degree of freedom allows one to view wave functionals as overlaps with Grassmann states, as is traditionally done for scalar fields.

Begin with the annihilation and creation operators for a bare, 2D system which satisfy

$$\{\hat{a}, \hat{a}^\dagger\} = \{\hat{b}, \hat{b}^\dagger\} = 1 \quad \text{while others vanish.} \tag{1}$$

The four available states are represented as $|0,0\rangle, |1,0\rangle, |0,1\rangle$ and $|1,1\rangle$; (2)

and considered orthonormal. The functionals representing states of the system are denoted as follows

$$|0,0\rangle = \Psi(0,0) \to 1, \ |1,0\rangle = \Psi(1,0) \to \bar{\eta}, \ |0,1\rangle = \Psi(0,1) \to \eta, \ |1,1\rangle = \Psi(1,1) \to \bar{\eta}\eta, \tag{3}$$

with the algebra $\hat{a}|0,0\rangle = \hat{b}|0,0\rangle = 0; \ \hat{a}^\dagger|0,0\rangle = |1,0\rangle; \ \hat{b}^\dagger|0,0\rangle = |0,1\rangle \ ; \hat{a}^\dagger\hat{b}^\dagger|0,0\rangle = |1,1\rangle.$ (4)

Thus overlaps are given by $\langle\bar{\eta}\eta|0,0\rangle = 1, \ \langle\bar{\eta}\eta|1,0\rangle = \bar{\eta}, \ \langle\bar{\eta}\eta|0,1\rangle = \eta, \ \langle\bar{\eta}\eta|1,1\rangle = \bar{\eta}\eta.$ Here the overbar means the Hermitian conjugate, used when we are speaking only of the algebra. When we move to space-time dependence, we will use the usual † symbol.

This implies $\langle \bar{\eta}, \eta | \bar{\eta}', \eta' \rangle = \sum_{n,m=0}^{1} \langle \bar{\eta}, \eta | n, m \rangle \langle n, m | \bar{\eta}', \eta' \rangle = 1 + \bar{\eta}\eta' - \bar{\eta}'\eta + \bar{\eta}\eta\bar{\eta}'\eta' = e^{\bar{\eta}\eta' - \bar{\eta}'\eta}$ (5)

An overlap of states is then written as $\langle \Psi_1 | \Psi_2 \rangle = \int d^2\eta' d^2\eta \langle \Psi_1 | \bar{\eta}\eta \rangle \langle \bar{\eta}\eta | \bar{\eta}'\eta' \rangle \langle \bar{\eta}'\eta' | \Psi_2 \rangle$

$$= \int d^2\eta' d^2\eta \langle \Psi_1 | \bar{\eta}\eta \rangle e^{\bar{\eta}\eta' - \bar{\eta}'\eta} \langle \bar{\eta}'\eta' | \Psi_2 \rangle$$

Moving to field theory in configuration space, introduce the Schrodinger field operators which satisfy

$$\{\hat{\psi}_\alpha(\mathbf{x}), \hat{\psi}_\beta^\dagger(\mathbf{y})\} = \delta_{\alpha\beta}\delta(\mathbf{x}-\mathbf{y}).$$ (6)

These can be represented as $\hat{\psi}_\alpha(\mathbf{x}) = \frac{1}{\sqrt{2}}\left(\eta_\alpha(\mathbf{x}) + \frac{\delta}{\delta\eta_\alpha^\dagger(\mathbf{x})}\right)$ and $\hat{\psi}_\alpha^\dagger(\mathbf{x}) = \frac{1}{\sqrt{2}}\left(\eta_\alpha^\dagger(\mathbf{x}) + \frac{\delta}{\delta\eta_\alpha(\mathbf{x})}\right)$ (7)

and act on wave functionals $\Psi(\eta^\dagger, \eta) = \langle \eta^\dagger, \eta | \Psi \rangle$. Letting $D^2\eta = D\eta^\dagger D\eta$ the inner product becomes

$$\langle \Psi_1 | \Psi_2 \rangle = \int D^2\eta' D^2\eta \langle \Psi_1 | \eta^\dagger\eta \rangle \langle \eta^\dagger\eta | \eta^{\dagger'}\eta' \rangle \langle \eta^{\dagger'}\eta' | \Psi_2 \rangle$$ (8)

with $\langle \eta^\dagger\eta | \eta^{\dagger'}\eta' \rangle = \exp\left[\int d^3x \eta_\alpha^\dagger(\mathbf{x})\eta_\alpha'(\mathbf{x}) - \eta_\alpha^{\dagger'}(\mathbf{x})\eta_\alpha(\mathbf{x})\right].$ (9)

One sometimes sees $\langle \eta^\dagger\eta | \eta^{\dagger'}\eta' \rangle = \exp\left[\int d^3x \eta_\alpha^\dagger(\mathbf{x})\eta_\alpha'(\mathbf{x}) + \eta_\alpha(\mathbf{x})\eta_\alpha^{\dagger'}(\mathbf{x})\right]$ (10)

which always serves to be confuse issues. If one uses the dual $\Psi^*$ to define the inner product by functional integration $\langle \Psi_1 | \Psi_2 \rangle = \int D^2\eta \langle \Psi_1 | \eta^\dagger\eta \rangle \langle \eta^\dagger\eta | \Psi_2 \rangle$ then the dual will be defined as

$$\Psi^*[\eta, \eta^\dagger] = \int D^2\eta' \exp(\eta'\eta^\dagger + \eta^{\dagger'}\eta) \bar{\Psi}[\eta', \eta^{\dagger'}].$$ (11)

For Gaussian states $\Psi = \exp(\eta^\dagger \Omega \eta)$, and these play the role of the vacuum in the theory, while $\Omega$ is called the covariance.

For the dual one has $\Psi^*[\eta, \eta^\dagger] = \int D^2\eta' \exp(\eta'\eta^\dagger + \eta^{\dagger'}\eta + \eta^{\dagger'}\Omega\eta')$ (12)

$$= \det(-\Omega^{-1}) \exp\left(\eta^\dagger (\Omega^\dagger)^{-1} \eta\right).$$ (13)

And finally for the inner product we find $\langle \Psi | \Psi \rangle = \int D^2\eta \Psi^*[\eta, \eta^\dagger] \Psi[\eta, \eta^\dagger] = \det(1 + \Omega^\dagger \Omega)$.

Now in the Boson theory, one defines the vacuum as the state that is annihilated by the operator $\hat{a}$. Applying the field trajectory equation to this object yields a differential equation, which one then solves to complete the construction. There are some subtleties in the Fermionic case which we feel can be made clearer with some preparatory remarks.

In this context we will compute the ground state of the non-interacting Fermion system. Consider the Hamiltonian $H = \int d^3x d^3y \psi^\dagger(\mathbf{x})h(\mathbf{x},\mathbf{y})\psi(\mathbf{y})$ with $h(\mathbf{x},\mathbf{y}) = [-i\gamma^0\gamma^i\partial_i + \gamma^0 m]\delta(\mathbf{x}-\mathbf{y})$ and $i=1,2,3$.

Let $\psi_n$ be the eigenmodes of the 1st quantized Hamiltonian $h$, as $h\psi_n = E_n\psi_n$. One can expand the $\eta_\alpha, \eta_\alpha^\dagger$ in terms of these eigenmodes, viz:
$$\eta_\alpha(\mathbf{x}) = \sum_n \eta_{\alpha,n}\psi_n(\mathbf{x}) \qquad \eta_\alpha^\dagger(\mathbf{x}) = \sum_n \eta_{\alpha,n}^\dagger \psi_n^\dagger(\mathbf{x}). \tag{14}$$

Apply the Hamiltonian $H$ to the Gaussian states (11). Here $\eta^\dagger \Omega \eta = \sum_{n,m,\alpha} \eta_{\alpha,n}^\dagger \Omega_{n,m} \eta_{\beta,m}$. Now put

$$\Omega(\mathbf{x},\mathbf{y}) = \sum_{n,m,\alpha} \psi_{\alpha,n}(\mathbf{x})\Omega_{n,m}\psi_{\beta,m}^\dagger(\mathbf{y}). \tag{15}$$

After some algebra, and utilizing the orthogonality properties of the eigenmodes, one gets for the ground state energy $E_0 = \frac{1}{2}\text{Tr}h(1+\Omega) = \frac{1}{2}\sum_n (1+\Omega_{nn})$. The elements of $\Omega$ are given as $\Omega_{nm} = \pm\delta_{nm}$. We get for equation (15) the result
$$\Omega(\mathbf{x},\mathbf{y}) = \sum_{E_n<0} \psi_{\alpha,m}(\mathbf{x})\psi_{\beta,n}^\dagger(\mathbf{y}) - \sum_{E_n>0} \psi_{\alpha,m}(\mathbf{x})\psi_{\beta,n}^\dagger(\mathbf{y}). \tag{16}$$

Nikolic [3] writes $\psi(\mathbf{x}) = \psi^{(P)}(\mathbf{x}) + \psi^{(A)}(\mathbf{x}) = \sum_k a_k u_k(\mathbf{x}) + \sum_k b_k^\dagger v_k(\mathbf{x})$, where $P$ refers to positive frequency components (Particles), $A$ refers to negative frequency components (anti-particles) and $u$ and $v$ represent the usual 4-spinors. In a box with sides of length L, one has (in this notation)

$$u(\mathbf{x}) = \frac{1}{\sqrt{V}}\sum_k u_{\mathbf{k},s} e^{i\mathbf{k}\cdot\mathbf{x}} \qquad v(\mathbf{x}) = \frac{1}{\sqrt{V}}\sum_k v_{\mathbf{k},s} e^{i\mathbf{k}\cdot\mathbf{x}} \quad \text{for} \quad k_i = \frac{2\pi n_i}{L} \quad i=1,2,3 \quad V=L^3 \tag{17}$$

and $n_i$ are positive or negative integers, including zero, and the spin index has been suppressed.

In the language of reference [3] one defines
$$\Omega(\mathbf{x},\mathbf{y}) = \Omega^A(\mathbf{x},\mathbf{y}) - \Omega^P(\mathbf{x},\mathbf{y}) = \sum_k v_k(\mathbf{x})v_k^\dagger(\mathbf{y}) - \sum_k u_k(\mathbf{x})u_k^\dagger(\mathbf{y}). \tag{18}$$

The vacuum is represented as $\Psi_0[\eta,\eta^\dagger] = N\exp\int d^3x \int dy^3 \eta^\dagger(\mathbf{x})\Omega(\mathbf{x},\mathbf{y})\eta(\mathbf{y})$. \hfill (19)

$N$ is a normalization constant, and the $\eta$'s are anti-commuting Grassmann numbers. We will now move to normal modes in a box of side $L$ and volume $V$. Excited states are formed by acting with the creation operators
$$\hat{a}_\mathbf{k}^\dagger = \frac{1}{V^{1/2}}\int_V d^3x \hat{\psi}^\dagger(\mathbf{x})u_\mathbf{k}(\mathbf{x}) \qquad \hat{b}_\mathbf{k}^\dagger = -\frac{1}{V^{1/2}}\int_V d^3x v_\mathbf{k}^\dagger(\mathbf{x})\hat{\psi}(\mathbf{x}) \tag{20}$$

on equation (19). The values for $\Omega_{nn}$ come from the demand $a_n^\dagger a_n \Psi = \begin{cases} 0 & \text{if } \Omega_{nn} = -1 \leftrightarrow E_n > 0 \\ 0 & \text{if } \Omega_{nn} = +1 \leftrightarrow E_n > 0 \end{cases}$, which selects a specific ground state. We can now get an explicit expression for $\Omega$. Noting that the square of $h_{(0)}^2 = p^2 + m^2$, consider [5] the vacuum energy

$$E_0 = \frac{1}{2}\text{Tr}h_{(0)}\Omega_{(0)} = \frac{1}{2}\sum_n E_n \Omega_{nn} = -\frac{1}{2}\sum_n |E_n|$$

$$= -\frac{1}{2}\text{Tr}\sqrt{p^2 + m^2} = -\frac{1}{2}\int_V d^3p\sqrt{p^2+m^2}.$$

NB, in the latter equation, the lattice sum uses the ansatz $\frac{1}{V}\sum_{\mathbf{k}} = \frac{1}{(2\pi)^3}\int_\infty d^3\mathbf{k}$. With all this we get

$$\Omega_{(0)} = \frac{1}{\sqrt{p^2+m^2}}\begin{pmatrix} m & \boldsymbol{\sigma}\cdot\mathbf{p} \\ \boldsymbol{\sigma}\cdot\mathbf{p} & -m \end{pmatrix} \text{ in the Dirac representation.}$$

## EMERGENCE OF PARTICLES

Let us begin with a simple observation to motivate the argument which follows. Write the free Dirac equation as $[-i\partial + m]_{ab}\psi_b = 0 = H^0_{ab}\psi_b$, where we have explicitly indicated the indices. Note that if we write $(H^0_{ca})^\dagger H^0_{ab}\psi_b = (H^0_{ca})^\dagger \delta^b_c H^0_{ab}\psi_b = H^2\psi_c = (\Box + m^2)\psi_c = 0$, then if there is an interaction term on the RHS $H^0_{ab}\psi_b = H^I_{ab}\psi_b$ and we multiply through by the same term, we get a field equation in terms of a nonlinear Klein Gordon equation $(\Box + m^2)\psi_c = (H^0_{ca})^\dagger H^I_{ab}\psi_b$. Writing a particular solution in terms of the Klein Gordon Green function yields $\psi_c(x) = \int G(x,y)[H^{0\dagger}_{ca} H^I_{ab}]\psi_b(y)dy$. Taking the Fourier Transform, it should come as no surprise that the integrand in the RHS contains the term $\frac{i\slashed{k}+m}{k^2+m^2}$; this is the usual way the solution to the Dirac equation with source is written. What it really tells us is that we have rewritten the equation in terms of a second order derivativation.

We will take the point of view that the field dynamics are to be fundamentally expressed in terms of the Lagrangian corresponding to the Klein Gordon equation. We may label these as observables, after a suitable re-labelling (bosonization). This is the point of view embraced in [3], and we will adopt it. From this point of view, all field components $\psi_a$ satisfy the Klein Gordon equation, $(\Box + m^2)\psi_a = 0$. The Dirac equation $[-i\partial + m]_{ab}\psi_b = 0$ is then regarded as a constraint, which leads to a definition of the Lorentz group to which the field $\psi_a$ belongs. Bosonization is necessary for at least two reasons: (1) observables cannot consistently be represented by Grassmann valued functions; (2) an attempt to find a positive definite-valued function in terms of the obvious candidates defined quadratically (e.g. the energy tensor in terms of $\bar{\psi}\psi$ and its derivatives) fails because expressions involving its commutator at different points involve derivatives of delta functions. In this language, the expression for the vacuum state can be written

$$N\exp[-\int \psi^\dagger(x)\sqrt{-\Delta+m^2}\psi(x)d^3x - iE_0 t] \text{ or as } N\exp[\int d^3x d^3y \psi^\dagger(\mathbf{x})h(\mathbf{x},\mathbf{y})\psi(\mathbf{y}) - iE_0 t] \qquad (21)$$

with $h(\mathbf{x},\mathbf{y})$ defined as above, i.e. the square root taken explicitly in the notation, while using a consistent definition for the operators in equation (7). We will assume this can be maintained, and write a general expression for $\psi$ in the polar form

$$\psi^a \equiv R e^{iS} \phi^a. \qquad (22)$$

This notation is from [6]. If $\psi$ is a scalar the superscript is unnecessary. If $\psi$ is a spinor, the phase is defined as follows: the scalar phase is that quantity that is common to all the $\phi^a$, the remaining expression is a spinor with scalar phase zero.

With the foregoing convention, in the non-interacting case, the machinery from Bohmian QFT may be invoked. To obtain an excited state for an "electron" apply the $\hat{a}^\dagger$ to the vacuum

$$\left\{ \frac{1}{\sqrt{V}} \int \frac{1}{\sqrt{2}} (\eta_a^\dagger + \frac{\delta}{\delta \eta_a}) u(\mathbf{x}) d^3 x \right\} N\exp\left[ \int d^3x d^3y \eta^\dagger(\mathbf{x})\Omega(x,y)\eta(\mathbf{y}) - iE_0 t \right] \qquad (23)$$

$$= \sqrt{\frac{2}{V}} u_\mathbf{k}^a \eta_{a\,\mathbf{k}}^\dagger N\exp[\cdots] \qquad (24)$$

Now one assumes the field trajectory equation $\frac{\partial \psi}{\partial t} = \frac{\delta S[\mathbf{x},t]}{\delta \psi}|_{\psi(\mathbf{x})=\psi(\mathbf{x},t)}$. An interesting question here is whether to utilize the phase of $\Psi e^{iS}$ evaluated *before* or *after* we Bosonize the field $\eta$. It is interesting that if one properly utilizes the commutation properties of the $\eta$ Grassmann variables, the result is the same. The algebra is simpler if one does the Bosonization first. To this end, introduce the following Transformation:

$$\eta \to q \quad \psi \to \varphi. \qquad (25)$$

This will result in the following:

$$|0,0\rangle = \Psi(0,0) \to 1, \ |1,0\rangle = \Psi(1,0) \to q^\dagger, \ |0,1\rangle = \Psi(0,1) \to q, \ |1,1\rangle = \Psi(1,1) \to q^\dagger q,$$

$$\langle q^\dagger q|0,0\rangle = 1, \ \langle q^\dagger q|1,0\rangle = q^\dagger, \ \langle q^\dagger q|0,1\rangle = q, \ \langle q^\dagger q|1,1\rangle = q^\dagger q. \qquad (26)$$

Traces are not the same, due to lack of anti-commutation, but we will not need them. We also have

$$\langle q^\dagger, q | q^{\dagger'}, q' \rangle = \sum_{n,m=0}^{1} \langle q^\dagger, q | n, m \rangle \langle n, m | q^{\dagger'}, q' \rangle = 1 + q^\dagger q' + q q^{\dagger'} + q^\dagger q' q q^{\dagger'} = e^{q^* q' + q q'^*} \qquad (27)$$

This implies that the dual $\quad \Psi^*[q, q^\dagger] = \int D^2 q' \exp(q' q^\dagger + q^{\dagger'} q) \overline{\Psi}[q^{\dagger'}, q'^\dagger] \qquad (28)$ just as before, and noting $\quad q^{\dagger'} q = q q^{\dagger'} \quad \eta'^\dagger \eta \neq \eta \eta'^\dagger$ .

Recalling equation (14) from reference [1], we will now have similarly

$$\Psi_0[q, q^\dagger, t] = N \prod_{\mathbf{k}} \exp(-\omega_{\mathbf{k}} q_{\mathbf{k}}^\dagger q_{\mathbf{k}} - i E_0 t), \qquad \omega_{\mathbf{k}}^2 = \mathbf{k}^2 + \mu^2 . \qquad (29)$$

The only difference is that the products (or sums if in an exponential) are not identical for $\pm \mathbf{k}$.

The trajectory equation yields $\quad \dot{q}_{\mathbf{k}} = \dfrac{\partial S}{\partial q_{\mathbf{k}}^\dagger} = \dfrac{1}{2 i q_{\mathbf{k}}^\dagger} \quad$ for each k, or $\quad \dot{\eta}_{\mathbf{k}} = \dfrac{\partial S}{\partial \eta_{\mathbf{k}}^\dagger} = \dfrac{1}{2 i \eta_{\mathbf{k}}^\dagger} \quad$ before the $\qquad (30)$ transformation. Now reference [3] pursues the Bosonization further, and expresses all field quantities explicitly in terms of real variables. This is due to the general nature of his guidance equations. That will not be necessary here, as we are only interested in the non-interacting case, with rectilinear motion at constant velocity.

As in [1], the solution of (30) is $q_{\mathbf{k}} = A e^{-i \omega_{\mathbf{k}} t}$ with $A = \dfrac{1}{\sqrt{2 \omega_{\mathbf{k}}}}$. Now, repeating the same argument, one

assumes we are in the particle's rest frame, where the scalar part is as in part I, and the spinor factor is

$u_{\mathbf{k}}^1 = 1$ for $|\mathbf{k}| = \mu = k$. We get then $\varphi = \dfrac{1}{\sqrt{V}} \sum_{\mathbf{k}} e^{i \mathbf{k} \cdot \mathbf{x}} \delta_{k,\mu} \int_0^t q_{\mathbf{k}} dt' = \dfrac{1}{\sqrt{V}} \sum_{\mathbf{k}} e^{i \mathbf{k} \cdot \mathbf{x}} \delta_{k,\mu} \int_0^t (q_{\mathbf{k}} + q_{\mathbf{k}}^*) dt'$ (31)

For the scalar part. The form of equation (31) is taken to be consistent with part I, except that $q_k \neq q_{-k}^\dagger$. This point is not entirely trivial, since it indicates how the various degrees of freedom are to be represented. Part I dealt with real fields, although with double degrees of freedom for convenience. Here we have a scalar phase associated with charge, and a spinor phase associated with spin, as assumed in equation (22). Since we are only dealing with free particles, the extraneous degree of freedom associated with charge can be dealt with by introducing $\varphi^a$ for $a = 1, 2, 3, 4$. Then we may associate $a = 1, 2$ for spin "up" and negative charge and $a = 3, 4$ for spin "down" and positive charge as in the usual case.

Then we may take equation (31)

to read $\quad \varphi^r = \dfrac{1}{\sqrt{V}} \sum_{\mathbf{k}} e^{i \mathbf{k} \cdot \mathbf{x}} \delta_{k,\mu} \int_0^t q_{\mathbf{k}} dt' = \dfrac{1}{\sqrt{V}} \sum_{\mathbf{k}} e^{i \mathbf{k} \cdot \mathbf{x}} \delta_{k,\mu} \int_0^t \dfrac{1}{2} (q_{\mathbf{k}} + q_{\mathbf{k}}^*) dt'$ (32)

or $\quad \varphi^r = \frac{1}{\sqrt{V}} \sum_{\mathbf{k}} e^{i\mathbf{k}\cdot\mathbf{x}} \delta_{k,\mu} \int_0^t q_{\mathbf{k}} dt' = \frac{1}{\sqrt{V}} \sum_{\mathbf{k}} e^{i\mathbf{k}\cdot\mathbf{x}} \delta_{k,\mu} \int_0^t \frac{1}{\sqrt{2\omega_k}} (\cos\omega_k) dt'$ (33)

$$= \frac{1}{\sqrt{V}} \sum_{\mathbf{k}} e^{i\mathbf{k}\cdot\mathbf{x}} \delta_{k,\mu} \int_0^t q_{\mathbf{k}} dt' = \frac{1}{\sqrt{V}} \sum_{\mathbf{k}} e^{i\mathbf{k}\cdot\mathbf{x}} \delta_{k,\mu} \theta(t) \frac{1}{\sqrt{2\omega_k}} \frac{\sin\omega_k t}{\omega_k}.$$ (34)

Finally, as in part I we get $\varphi^a = \frac{1}{\sqrt{2V\omega_\mu}} \frac{\mu}{2\pi^2} \frac{\sin(\mu r)}{r} \theta(t) \frac{\sin(\omega_\mu t)}{\omega_\mu}$ where $r = \sqrt{x^2 + y^2 + z^2}$. (35)

Picking a choice for $a$, we will define $\quad \varphi = \varphi^a u^a_{k=\mu} \quad$ for $a = 1$. (36)

As a reminder, in this notation $u^1 = \begin{bmatrix} 1 \\ 0 \\ 0 \\ 0 \end{bmatrix}$ $u^2 = \begin{bmatrix} 0 \\ 1 \\ 0 \\ 0 \end{bmatrix}$ $u^3 = \begin{bmatrix} 0 \\ 0 \\ 1 \\ 0 \end{bmatrix}$ $u^4 = \begin{bmatrix} 0 \\ 0 \\ 0 \\ 1 \end{bmatrix}$ in the rest frame.

And since $\frac{\mu}{\omega_\mu} = 1$ in the rest frame we finally get $\varphi = \frac{u^1_{k=\mu}}{\sqrt{2V\omega_\mu}} \frac{1}{2\pi^2} \frac{\sin(\mu r)}{r} \theta(t) \sin(\omega_\mu t)$. (37)

Now one introduces a Lorentz boost with velocity $v$ in the z direction. Spinors are transformed with the operator $S = \exp\left[-\frac{\omega}{2} \frac{\boldsymbol{\alpha}\cdot\mathbf{v}}{|\mathbf{v}|}\right]$ acting on the matrix whose columns are formed by the $u^i$ [8]. Here $\omega = \tanh\left(\frac{-|\mathbf{v}|}{c}\right)$ and $\alpha_z = \sigma_3$, the Pauli matrix. The final result is to no one's surprise

$$\varphi = \frac{1}{\sqrt{2V\omega_\mu}} \frac{1}{2\pi^2} \sqrt{\frac{E+m}{2m}} \begin{bmatrix} 1 \\ 0 \\ \frac{p}{E+m} \\ 0 \end{bmatrix} \frac{\sin(\mu\sqrt{\rho^2 + \gamma(z-vt)^2}}{\sqrt{\rho^2 + \gamma(z-vt)^2}} \theta(\gamma(ct - v/cz)) \sin(\omega_\mu \gamma(ct - v/cz)).$$ (38)

**DISCUSSION**

The first point to observe is that we have considered only the simplest case, a non-interacting particle moving in rectilinear motion at constant velocity. Even in this case, some issues have been only lightly discussed. In the equation $\psi_c(x) = \int G(x,y)[H^{0\dagger}_{ca} H^I_{ab}]\psi_b(y) dy$ the RHS makes clear there is component mixing if there are any interaction terms to be considered. So the notion of "Particle" will be dependent upon context. What interactions or boundaries (which are also significant interactions) should be considered? Of course this is old news in quantum field theory, and the issue becomes nearly obliterated in curved spacetime. Our subject is speculative physics, and so mathematical considerations take a back seat; but issues of superfluous degrees of freedom should be kept in mind. In reference [1] we considered a complex formulation of a real field, which was really two oscillators to represent the field. We have done the same here, and doubled the degrees of freedom again to allow for spin. These all have a natural interpretation, but it would be interesting to see if the issue of particles emerging from the field trajectory equation (30) depends critically upon these extra degrees. This point is meant to be reflected in the title, "How particles *can* emerge…" and not "How particles *must* emerge…".

Much formalism was introduced and seemingly not applied to the main argument. This was done to address natural objections that immediately arise when one is unfamiliar with the subtleties of quantum field theory in the Schrodinger picture, as was the author. The necessity of Bosonization is manifest, but the real utility of this procedure only becomes clear when one considers a formulation that allows for interactions. This was done in [3] and [7]. When this is done, a host of fundamental questions of principle emerge: primary among these is which equation to use for particle trajectories. Nikolic points out that in Fermionic field theory, both the Dirac and Klein Gordon currents are conserved; and this fact allows for a consistent formulation of theory for general statistics.